\documentclass{emulateapj}

\shorttitle{Breakdown of K-S law at GMC scales in M33}
\shortauthors{Onodera et al.}
\begin{document}
\newcommand{\mo}{\: M_\odot}
\newcommand{\pc}{\: \rm pc} 
\newcommand{\kpc}{\:\, \rm kpc}
\newcommand{\ghz}{\: \rm GHz}
\newcommand{\mhz}{\: \rm MHz}
\newcommand{\kms}{\: \rm km\,s^{-1}}
\newcommand{\yr}{\:\rm yr}
\newcommand{\Jy}{\:\rm Jy}
\newcommand{\K}{\:\rm K}
\newcommand{\mK}{\:\rm mK}
\newcommand{\cm}{\:\rm cm}

\renewcommand{\micron}{\ensuremath{\mu\mathrm{m}\ }}

\title{Breakdown of Kennicutt-Schmidt Law at GMC Scales in M33}
\author{Sachiko~Onodera\altaffilmark{1,2}, Nario~Kuno\altaffilmark{1, 3},
Tomoka~Tosaki\altaffilmark{1, 4}, Kotaro~Kohno\altaffilmark{2},
 Kouichiro~Nakanishi\altaffilmark{3, 5},
 Tsuyoshi~Sawada\altaffilmark{5}, Kazuyuki~Muraoka\altaffilmark{6}, Shinya~Komugi\altaffilmark{7}, Rie~Miura\altaffilmark{8}, Hiroyuki~Kaneko\altaffilmark{3}, Akihiko~Hirota\altaffilmark{1}, \\ and \\ Ryohei~Kawabe\altaffilmark{1}}
\altaffiltext{1}{Nobeyama Radio Observatory, National Astronomical
Observatory, 462-2, Nobeyama, Minamimaki, Minamisaku, Nagano 384-1305, Japan}
\altaffiltext{2}{Institute of Astronomy, The University of Tokyo,
2-21-1 Osawa, Mitaka, Tokyo 181-0015, Japan}
\altaffiltext{3}{The Graduate University for Advanced Studies (Sokendai), 2-21-1 Osawa, Mitaka, Tokyo 181-8588, Japan}
\altaffiltext{4}{Joetsu University of Education, Yamayashiki-machi,
Joetsu, Niigata 943-8512, Japan}
\altaffiltext{5}{ALMA Project Office, National Astronomical Observatory,
2-21-1 Osawa, Mitaka, Tokyo 181-8588, Japan}
\altaffiltext{6}{Osaka Prefecture University, 1-1 Gakuen-cho, Nakaku, Sakai, Osaka 599-8531, Japan}
\altaffiltext{7}{Japan Aerospace Exploration Agency,
Institute of Space and Astronautical Science
3-1-1 Yoshinodai, Sagamihara, Kanagawa 229-8510, Japan}
\altaffiltext{8}{Department of Astronomy, Graduate School of Science, The University of Tokyo, 7-3-1 Hongo, Bunkyo-ku, Tokyo 113-0033, Japan}

\email{sonodera@nro.nao.ac.jp}

\begin{abstract}
We have mapped the northern area ($30'\times 20'$) of a local group spiral galaxy M33 in $^{12}$CO({\it J}\,=\,1--0) line with the 45-m telescope at the Nobeyama Radio Observatory.
Along with H$\alpha$ and Spitzer 24-\micron data, we have investigated the relationship between the surface density of molecular gas mass
 and that of star formation rate (SFR) in an external galaxy (Kennicutt-Schmidt
 law) with the highest spatial
 resolution ($\sim\!80\pc$) to date, which is comparable to scales of giant molecular clouds (GMCs). At positions where CO is significantly detected, the SFR surface density exhibits a wide
 range of over four orders of magnitude, from $\Sigma_{\rm
      SFR}\lesssim10^{-10}$ to $\sim\!10^{-6}\mo\yr^{-1}\,\pc^{-2}$, whereas the
 $\Sigma_{\rm H_2}$ values are mostly within $10\textendash40\mo\pc^{-2}$. The surface density of gas and that of SFR correlate well at a $\sim$1-kpc resolution, but the correlation becomes looser with higher resolution and breaks down at GMC scales. The scatter of the $\Sigma_{\rm SFR}$--$\Sigma_{\rm H_2}$ relationship in the $\sim\!80$-pc resolution results from the variety of star forming activity among GMCs, which is attributed to the various evolutionary stages of GMCs and to the drift of young clusters from their parent GMCs. This result shows that the Kennicutt-Schmidt law is valid only in scales larger than that of GMCs, when we average the spatial offset between GMCs and star forming regions, and their various evolutionary stages.
 
\end{abstract}

\section{INTRODUCTION}

Understanding the formation process of massive stars in galaxies is
one of the key issues in modern astronomy because massive stars play
essential roles in the evolution of galaxies. Because stars are formed
from molecular gas, the study of molecular gas in galaxies 
provides us with important clues to the fundamental physical
processes of massive star formation in galaxies. 
Observational studies of galaxies on global scales have shown that the surface density of star formation rate (SFR) and that of cold gas obey a simple relation,
$\Sigma_{\rm SFR}\propto\Sigma_{\rm gas}^n$, where $\Sigma_{\rm SFR}$ is the SFR per unit area and $\Sigma_{\rm gas}$ is the surface density of gas. This relationship is known as the Kennicutt-Schmidt law \citep[hereafter, K-S law:][]{schmidt59, kennicutt98}. 
It has been shown that the disk-averaged surface density of SFR is much
better correlated  
with that of molecular gas $\Sigma_{\rm H_2}$ than with that of total (H{\sc i}\,+\,H$_2$) gas $\Sigma_{\rm HI + H_2}$ \citep[e.g.,][]{wong02}. 

Because almost all of the existing studies on the K-S law have been carried out based on CO observations of kpc-scale resolution or disk-averaged data \citep[e.g.,][]{komugi05},
 not much is known about the validity of the K-S law for smaller molecular structures such as giant molecular associations (GMAs) and giant molecular clouds (GMCs). Recently, \citet{kennicutt07} investigated K-S law in M51 down to a 
linear scale of 500\,pc and cloud mass scales of $10^6\textendash10^7\mo$. 
They found that the nonlinear K-S law can be extended down to that scale.
\citet{bigiel08} revealed that the K-S law is applicable at the 750-pc scale for seven spiral galaxies, and that it holds down to the 250-pc scale in M51. \citet{verley10} have shown that a loose correlation exists even in the 180-pc scale in M33, where a strong correlation was found on the global scale \citep{heyer04}.
Because most of the molecular gas is confined within molecular clouds and virtually all of the GMCs are sites of star formation in the Milky Way, GMCs play an important role in star formation. It is important to understand how star formation at GMC scales is linked to the K-S law, which is valid for kpc-scales.

One way to address this issue is to conduct a high-spatial-resolution mapping of entire
molecular gas disks in the nearest galaxies. 
The recent improvements in the resolution, sensitivity, 
and observation efficiency of radio telescopes have made such observational studies feasible.

M33 is the best target for this purpose. It is one of the nearest spiral galaxies 
\citep[$\rm D=840$ kpc;][]{freedman91} in which individual GMCs can be resolved using present-day 
instruments. Furthermore, because its disk is moderately face-on ($i=51^\circ$), it is easy to study the molecular gas in relation to star formation.

Because of these desirable characteristics, M33 has been the subject of many large-scale studies of GMCs. 
Pioneering surveys of the central 1.7-kpc region of M33 were conducted 
with the 12-m telescope at the National Radio Astronomy Observatory \citep{ws89} and 	
the Owens Valley Radio Observatory interferometer \citep{ws90}. They revealed that the properties of GMCs (velocity width, diameter, peak brightness temperature, and mass) were very similar to those in the Milky Way Galaxy. 
More recently, a CO\,({\it J}\,=\,1--0) all-disk survey was performed with the Berkeley-Illinois-Maryland Association (BIMA) \citep{BIMAI}, 
which identified 148 GMCs with masses over $1.5\times 10^5\mo$ across the star-forming disk. 
The BIMA survey recovered only 20\% of the CO flux in M33 
because of the missing short spacings of the interferometer \citep{heyer04}; thus, \cite{BIMAIII} performed 
additional CO\,({\it J}\,=\,1--0) observations with the 45-m telescope at the Nobeyama Radio Observatory (NRO), 
and combined them with the existing data obtained from BIMA and the 14-m telescope at the Five College Radio Astronomy Observatory (FCRAO). 
They obtained the CO\,({\it J}\,=\,1--0) map in the inner $\sim\!15'\times 15'\, (\sim\!3.7\kpc)$ region 
at a resolution of $20''\,(\sim\!80\pc)$. 

However, these observations are still insufficient to understand the overall nature of the GMCs in M33 because they do not cover the entire disk, and they possess inhomogeneities due to multi-field interferometric observations.
Accordingly, a homogeneous wide-area single-dish mapping with a high spatial resolution and high sensitivity is 
required in order to determine the molecular gas properties of the entire disk. 
In order to achieve this purpose, it is effective to adopt the on-the-fly (OTF) mapping technique using a single-dish telescope having a large aperture.
We performed an M33 all-disk $^{12}$CO({\it J}\,=\,1--0) survey with the 45-m telescope, as an NRO legacy project to study the properties and evolution of GMCs and star
formation within the entire galaxy. In this paper, we investigate the K-S law in the GMC scale, which is the highest resolution ($\sim\!80\pc$) to date, as an initial result of this project.

\section{OBSERVATIONS AND DATA}
We observed the $^{12}$CO\,({\it J}\,=\,1--0) lines toward the north in a $30'\times 20'$ ($\rm\sim\!7.3\kpc\times 4.9\kpc$)
region of M33;
the observation region covered the galactic center and the three major H{\sc ii} regions: 
NGC\,604, NGC\,595, and IC\,133. The mapping area is shown in
Fig.~1.  

The observations were carried out from 
January to April, 2008, with the 45-m telescope at NRO.
 The beam size of the telescope (HPBW) was $15''$ for the rest frequency of the
$^{12}$CO\,({\it J}\,=\,1--0) line (115.271204 GHz). We used the 25-BEam Array Receiver System
\citep[BEARS:][]{sunada00}. 
Because BEARS is a double sideband receiver, we measured the sideband ratios of each beam of BEARS by observing a bright standard source NGC 7538 using both BEARS and a single-beam receiver S100 equipped with a single sideband filter.
The errors in the scaling factors of BEARS were smaller than $\sim\!\!20\%$. 
The main beam efficiency measured with S100 was $\eta_{\rm MB}=0.32\pm
0.02$ at 115 GHz.  
The typical system temperature was $\sim\!700\K$ (SSB). 
We used digital autocorrelators as backends. 
The spectrometers have 1024 channels covering a bandwidth of $512 \mhz$ with
a frequency resolution of 1.0 MHz, which corresponds to $1332\kms$ and $2.6\kms$. 
The chopper-wheel technique was employed to calibrate the antenna
temperature $T_{\rm A}^*$. Hereafter, all the CO intensity measurements are mentioned in the $T_{\rm MB}$ ($\equiv\! T_{\rm A}^*/\eta_{\rm MB}$) scale. 

The observations were performed using the OTF observation mode
\citep{sawada08}.
Throughout the observations, the data were
sampled every 0.1 s. The emission-free ``OFF" points were taken at an offset of $30'$ from the center. We checked the pointing accuracy every hour with a 5-point
observation of a SiO maser source IRC+30021 with a $43\,\ghz$ SIS receiver
(S40). 
We excluded all the data in the observation sequence that had
pointing errors larger than $7''.5$ or wind
speed  $V>10\: \rm m\,s^{-1}$ to avoid systematic intensity
loss due to pointing errors. After these data screenings, the net integration time was determined to be approximately 130 h.  

The data reductions were made with the OTF reduction software package NOSTAR, which was implemented by the
NRO. The data were convolved with a Gaussian-tapered Jinc function to
create the spectral data cube. The grid spacing was taken as $7''.5$,
and the spatial resolution of the final map was $19''.3$. 
Finally, the baselines were fitted and subtracted from the data cube. The velocity resolution of the final cube is
$2.5\kms$. The rms noise in a velocity channel was $130\mK$. The rms noise in the integrated intensity map was $1.6\K\kms$, which corresponds to a mass sensitivity of $4.8\mo\pc^{-2}$.

\section{RESULTS}
\subsection{CO\,({\it J}\,=\,1--0) Map}
Fig.~1 shows the CO\,({\it J}\,=\,1--0) integrated intensity map
of our observation field for M33. It reveals the clumpy nature of the molecular gas distribution. 
The size of these clumps ranges from $\lesssim\!100\pc$ (comparable to our beam size) to $500\pc$, which corresponds to that of GMCs and/or GMAs. 
The northern arm that can be traced 
from ($\alpha,\,\delta$) = ($\rm 1^h33^m40^s$, $+30^\circ46'0''$) 
to ($\rm 1^h34^m30^s$, $+30^\circ49'0''$) is prominent.

The surface mass density of molecular hydrogen was estimated 
by applying the $^{12}$CO-to-H$_2$ conversion factor, 
$X_{\rm CO}=3\times 10^{20} \cm^{-2}\,\rm(K\kms)^{-1}$ \citep{ws90} as follows:

\begin{eqnarray}
\left( \frac{\Sigma_{\rm H_2}}{\mo\pc^{-2}}\right)
 &=& 4.81 \left(\frac{X_{\rm CO}}{3\times 10^{20}\cm^{-2}\,(\rm K\kms)^{-1}}\right) \nonumber\\
&   &    \times\left(\frac{I_{\rm CO}}{\K\kms}\right)\cos i,
\end{eqnarray}
where {\it i} is the inclination angle of the galaxy ($i=51^\circ$). 

\subsection{Star Formation Rate}
We calculated the SFR using the relation between the extinction-corrected H$\alpha$ line emission and the SFR, as presented in \citet{calzetti07}:
\begin{eqnarray}
\rm SFR(\!{\it \mo}\, yr^{-1})&=&5.3\times 10^{-42}\,
 [L(H\alpha)_{obs}  \\
& &      +(0.031\pm 0.006)L(24\,\mu\rm m)], \nonumber
\end{eqnarray}
in turn leading to
\begin{eqnarray}
\Sigma_{\rm SFR}(\!{\it \mo}\rm\,yr^{-1}\,pc^{-2})&=&5.0\times 10^{-5}\,
 [\Sigma(\rm H\alpha)_{obs} \\
&+&(0.031\pm 0.006)\Sigma(24\,\mu{\rm m})]\cos {\it i}, \nonumber
\end{eqnarray}
where the luminosities are in erg\,s$^{-1}$ and L\,(24\,$\mu$m) is expressed as $\nu\, L(\nu)$. 
The unit of $\Sigma(\rm H\alpha)_{obs}$ and $\Sigma(24\,\mu{\rm m})$ is $\rm erg\, s^{-1}\, cm^{-2}$.

We use the H$\alpha$ image of M33 as given by \citet{hoopes00}. It was
obtained using the 0.6-m Burrell-Schmidt telescope at the Kitt Peak National
Observatory. The dimensions of the CCD are $2048 \times 2048$ with
a pixel size of $2''.03$ and the total field of view is approximately $70'\times
70'$. The sensitivity is $0.8 \times 10^{-17}\, \rm erg\, s^{-1}\,
cm^{-2}\, arcsec^{-2}$ \citep{hoopes01}. 

We retrieved the 24-\micron Multiband Imaging Photometer 
\citep[MIPS:][]{rieke04} datasets (AORs 3647744, 3648256,
3648512, 3648768) of the basic calibrated data (BCD) 
created by the Spitzer Science Center (SSC) pipeline from the Spitzer Space Telescope 
\citep{werner04} data archive.
The mosaics were assembled by gathering all the BCDs except the first data
frames of each observation because they have shorter integration times. The individual
calibrated frames were processed using MOPEX \citep{makovoz05} for cosmic-ray rejection
and background matching was applied to overlapping fields of
view. The final dimension of the each mosaic was $21'
\times 89'$. After we removed the zodiacal light, and the noisy pixels near the
edges by trimming, the four images were aligned with their coordinates and combined. 
The FWHM of point spread function is $5''.7$.

The H$\alpha$ and 24-\micron images were convolved into the same angular resolution as
 the CO({\it J}\,=\,1--0) data ($19''.3$),
regridded to $7''.5$ per pixel, and trimmed to match the
CO\,({\it J}\,=\,1--0) map.  
The rms noise is $5\times\rm 10^{-7}\,erg\,s^{-1}\,cm^{-2}$ in the H$\alpha$ map, 
and $3\times\rm 10^{-5}\,erg\,s^{-1}\,cm^{-2}$ in the 24-\micron map. 
The resultant SFR image is shown in Fig.~2 with the overlaid CO\,({\it J}\,=\,1--0) contours. 
The rms noises in the H$\alpha$ and 24-\micron maps result in a $\Sigma_{\rm SFR}$ error 
and a sensitivity limit of $5\times 10^{-11}{\it \mo}\,\rm \,yr^{-1}\,pc^{-2}$. 

Fig.~2 reveals a striking variety of star formation properties. In regions where CO emission is detected, there is a large dispersion in the values of $\Sigma_{\rm SFR}$ of over four orders of magnitude. For example, the most massive GMC at ($\alpha$, $\delta$) = ($\rm 1^h33^m9^s.6, +30^\circ49'7''.3$) has little star formation activity despite the large amount of
molecular gas it contains.
In the three major star-forming regions (NGC\,604, NGC\,595, and IC\,133), there is a considerable difference in the amount of molecular gas associated with these star-forming regions. 
NGC\,604 has its associated GMCs at the same position, NGC\,595 has the ones
offset from it, and moreover, none of the GMCs are associated with IC\,133.  

\section{DISCUSSION}
Fig.~3 presents the relationship between the SFR and molecular hydrogen surface densities for four angular resolutions: $19''.3$ ($\sim\!80\pc$), $60''$ ($\sim\!240\pc$), $120''$ ($\sim\!500\pc$), and $240''$ ($\sim\!1\kpc$).
The panel of $\sim\!80\pc$ resolution is the K-S law plot with the highest spatial resolution for an extra-galaxy to date. Fig.~3 shows that apparent correlations exist between $\Sigma_{\rm H_2}$ and $\Sigma_{\rm SFR}$ at the $\sim\!1\kpc$ resolution, as already found in the disk-averaged data of M33 \citep{heyer04}. 
The best least-squares fit for the $\sim\!1\kpc$ resolution data with $\Sigma_{\rm H_2}>2\sigma$ is
\begin{equation}
 \log \Sigma_{\rm SFR}=(1.18\pm 0.11)\log\Sigma_{\rm H_2}-(9.38\pm0.05).
\end{equation} We have fitted the power law only to the data at the 1-kpc resolution, because the data at the other resolutions have a significant selection bias due to the omission of data points having $\rm \Sigma_{H_2}<2\sigma$. The selection bias makes the correlation slopes appear steeper than real.

Fig.~3 also shows that the correlation becomes looser with higher resolution, and it is hardly visible in the plot for $\sim\!80\pc$ resolution.  
Although most $\Sigma_{\rm H_2}$ values lie within $10\textendash40\mo\pc^{-2}$ 
where $\rm \Sigma_{H_2}>2\sigma$, $\Sigma_{\rm SFR}$ values exhibit a wide range of over four orders 
of magnitude, from $\lesssim\!10^{-10}$ to $\sim\!10^{-6}\mo\yr^{-1}\pc^{-2}$, 
indicating that the K-S law becomes invalid at this resolution.
That is, the K-S law is valid only for averaging SFR and gas mass in large scales of several hundred parsecs.  
Our resolution, $\sim\!80\pc$ is smaller than these scales.

We have examined the effects of stochasticity on the estimatiion of SFR due to small sampling at high 
spatial resolutions. The initial mass function is not fully populated when smaller regions that 
contain only a few stars in clusters are sampled. Thus, in regions with weaker extinction-corrected 
H$\alpha$ emission, this effect may lead to significant scatter in the estimated $\Sigma_{\rm SFR}$. 
We have estimated the significance of this effect on the 80-pc resolution data. For instance, 
$\Sigma_{\rm SFR}=10^{-9}\mo\yr^{-1}\pc^{-2}$ corresponds to the total flux of extinction-corrected
 H$\alpha$ emission having $L(\rm H\alpha)_{corr}=6.6\times 10^{35} erg\, s^{-1}$
 within the 80-pc beam. Assuming a gas temperature of $10^4\K$ and an electron density of 
$100\rm\, cm^{-3}$ in H{\sc ii} regions, 
the ionizing photon flux is given by $Q(\rm H^0)=7.3\times 10^{11} L(\rm H\alpha)_{corr}\,s^{-1}$ 
\citep{kennicutt88,brocklehurst71}. 
Therefore, the corresponding total ionizing photon flux is $Q(\rm H^0)=4.9\times10^{48} \rm s^{-1}$. 
According to Fig.~3 in \cite{cervino02}, $Q(\rm H^0)$ per unit mass varies with cluster age from 
$10^{47}$ to $10^{44}\, {\rm s}^{-1} \mo ^{-1}$. The ratio of the error in $Q(\rm H^0)$ to the mean 
value, $\sigma [Q({\rm H^0})]/Q({\rm H^0})$, is $\sim 10 M^{-1/2}$ (the ratio can vary by a factor 
$\sim 2$ depending on the cluster age), where $M$ is the cluster mass. Thus, the cluster mass required 
to produce the ionizing flux mentioned above is estimated to be 50 to $5\times10^4 \mo$, which 
results in $\sigma[Q({\rm H^0})]/Q({\rm H^0})\approx 1$ to 0.04. In summary, in the 80-pc scale, the 
error due to the effect of stochasticity at 
$\Sigma_{\rm SFR}=10^{-9}\mo\yr^{-1}\pc^{-2}$ is $\approx 0.04\textendash 1\times 10^{-9}\mo\yr^{-1}\pc^{-2}$. The error decreases with larger 
$\Sigma_{\rm SFR}$ and lower spatial resolutions. For example, the maximam $\sigma [Q({\rm H^0})]/Q({\rm H^0})$ in the 80-pc scale is 
 1, 0.4, and 0.1, and in the 240-pc scale it is 0.5, 0.1, and 0.05 for $\Sigma_{\rm SFR}=10^{-9}, 10^{-8}$, 
 and $10^{-7} \mo\yr^{-1}\pc^{-2}$, respectively. Although the error is significant at smaller $\Sigma_{\rm SFR}$ at the 80-pc resolution, it cannot explain the scatter of over four orders of magnitude in $\Sigma_{\rm SFR}$. 

Now, we consider the possible causes for the breakdown of the K-S law at high spatial 
resolutions. In grand design spiral galaxies, we often see a systematic offset between the molecular 
gas arms and the star forming arms as a result of the density wave and the time delay between the 
accumulation of gas and its ionization by newborn stars \citep[e.g.,][]{egusa09}. If such an offset 
exists in M33, it may result in the breakdown of the K-S law when we observe the galaxy with a spatial 
resolution comparable to the offset.
However, there is no such systematic offset between the CO and the star-forming arm in M33, 
as shown in Fig.~2. Therefore, the offset can be ruled out as a cause. 

In the Milky Way, young clusters recede from their parent GMCs with velocities of at least
$\sim\!10\kms$ \citep{leisawitz89}. As they evolve, the separation between clusters and their parent GMCs becomes significant and is
expected to be $\sim\!100\pc$ at a cluster age of $\sim\!10\,\rm Myr$. This drift scale is comparable to our resolution, $\sim\!80\pc$, and thus, the separation could well lead to the scatter in our plot. However, the existence of H{\sc ii} regions without neighboring GMCs and GMCs without neighboring H{\sc ii} regions in M33, such as the ones mentioned in the previous section, cannot be explained by drift alone. 

The other cause is the difference in evolutionary stages. As we have shown in Fig.~2, 
there is a large diversity in the star formation activity among GMCs. 
Some GMCs in the Milky Way are at least partially consumed and dissociated by newborn clusters
 \citep{leisawitz89}. Such GMCs are shown in Fig.~2 (i.e., the GMC at ($\alpha$, $\delta$) = 
($\rm 1^h34^m17^s$, $\rm 30^d52^m00^s$)). The diversity among GMCs arises from the evolutionary 
stages that each GMC goes through, as has been shown in the most active star forming region NGC\,604 
\citep{miura10}. \cite{chen10} studied the relation between $\Sigma_{\rm SFR}$ and 
$\Sigma_{\rm HI + H_2}$ in LMC for six GMCs in different stages of evolution, which are determined 
from the association of the underlying stellar population. They also found a large scatter in 
$\Sigma_{\rm SFR}$, which exhibits a range of two orders of magnitude whereas the $\Sigma_{\rm HI + H_2}$ 
lies within a factor of two.
\citet{kawamura09} classified molecular clouds in the LMC into three types
according to the activities of massive star formation: those with no signature of
massive star formation, those with relatively small H{\sc ii} region(s), and those with both H{\sc ii} region(s) and young stellar cluster(s). 
They concluded that there is a sequential relation between these three types. That is, molecular 
clouds are first formed; then, stars are formed in the molecular clouds; and finally, the newborn 
stars consume and dissociate the surrounding gas. This scenario is also applicable to M33, as can be
 shown in NGC\,604 by comparing it with warm/dense gas traced by CO\,({\it J}\,=\,3--2) lines \citep{tosaki07}.

Note that the extinction-corrected H$\alpha$ emission is not really an appropriate tracer of SFR at
resolutions as high as $\sim\!80\pc$, because there is a spatial displacement between ionizing stars 
and the surrounding H{\sc ii} gas. According to \citet{relano09}, the displacement is $\sim\!10\pc$ 
to $50\pc$ in M33, and there is a radial gradient of emission from the center to the boundary of
H{\sc ii} regions; in the order of ultraviolet, H$\alpha$, 24 $\mu$m, and CO emission. Because of
the radial gradient, the separations between ionizing stars and CO are larger than that between 
H$\alpha$ and CO. Thus, the ``real" SFR would produce an even larger scatter in the K-S law plot 
for the $\sim$80-pc resolution.
The contribution of dust heating from older stars to the 24-\micron emission can also be responsible for the scatter, although its effect is minor.

Owing to our spatial resolution, which is comparable in scale to that of GMCs, 
we have clearly revealed the variety in the evolutionary stages among GMCs. 
The different evolutionary stages of GMCs are the main cause of deviations from the
K-S law. The higher the spatial resolution, the more individual star-forming regions
and molecular clouds we can resolve, and the larger the deviations seen in the K-S law plots.

\section{SUMMARY} 
We conducted on-the-fly mapping of the northern part of M33 ($30'\times 20'$)
in $^{12}$CO\,({\it J}\,=\,1--0) lines with the 45-m telescope at the Nobeyama Radio Observatory (NRO).  
We tested the relationship between the surface density of molecular gas mass and the SFR in an external galaxy with the
highest resolution ($\sim\!80\pc$) to date. We found that the star-forming activities in the GMC scale
 exhibit a wide range of over four orders of magnitude at positions where CO is significantly 
detected, such that $\Sigma_{\rm SFR}\lesssim\!10^{-10}$ to $\sim\!10^{-6}\mo\yr^{-1}\,\pc^{-2}$, 
whereas their $\Sigma_{\rm H_2}$ values are mostly in the range of $10\textendash40\mo\pc^{-2}$. 
We also found that the K-S law of H$_2$ gas, $\Sigma_{\rm SFR}\propto\Sigma_{\rm H_2}^n$, 
becomes invalid at GMC scales ($\sim\!80\pc$). The breakdown of the K-S law is attributed to the 
difference in the evolutionary stages of GMCs and the drift of newborn clusters from their parent GMCs
that become apparent at this spatial resolution.

\acknowledgements
We thank the referee, Robert Kennicutt, for his prompt report and for his helpful comments, which improved our work.
We also thank Rene Walterbos for providing us with the H$\alpha$ image
of M33. S. O. was financially supported by the Global COE Program ``The
Physical Sciences Frontier," MEXT, Japan. The Nobeyama Radio Observatory is a branch of the National Astronomical Observatory of Japan, National Institutes of Natural Sciences.

\begin{figure}[p]
\begin{center}
\plotone{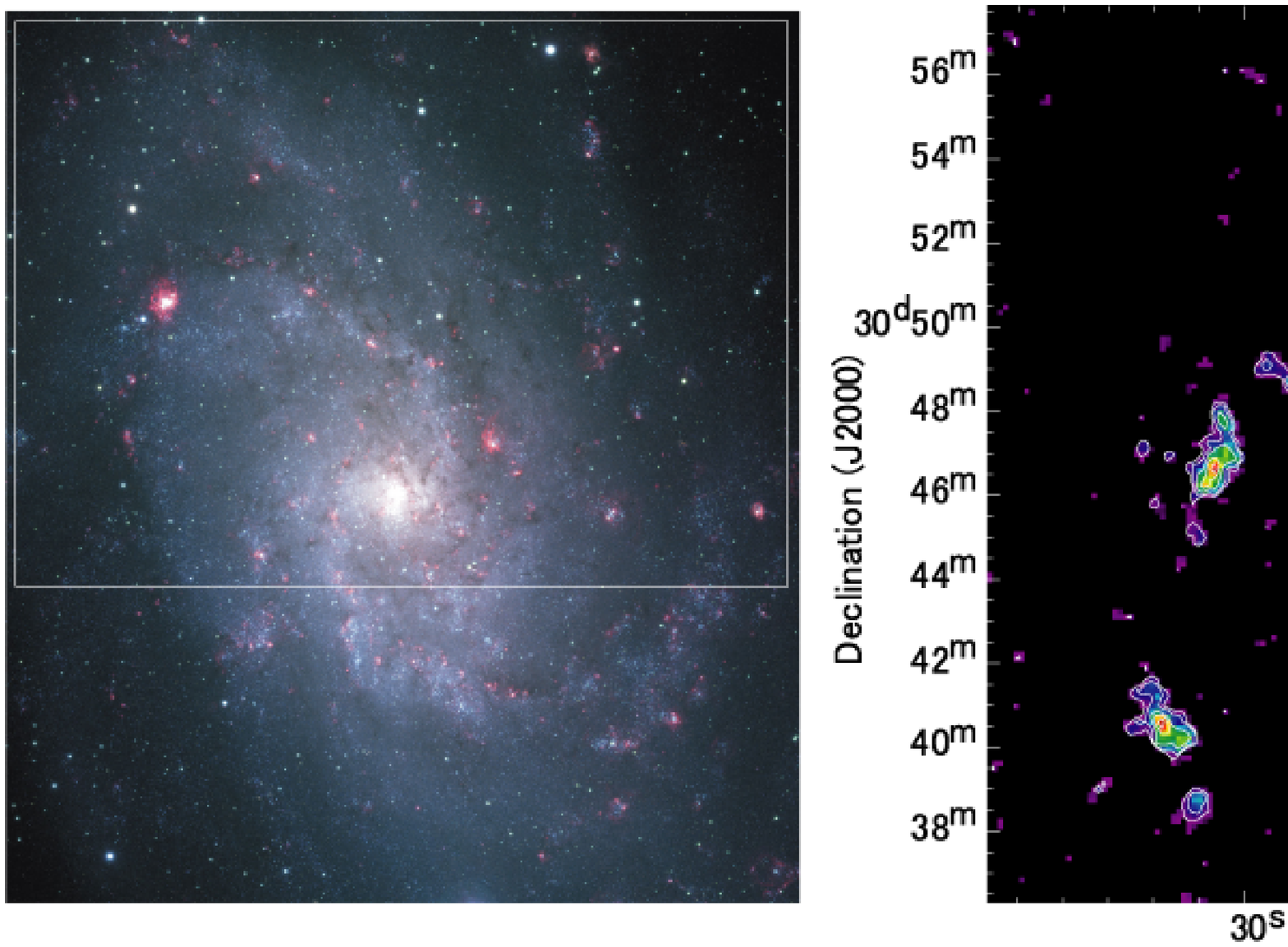}
\end{center}
\caption{(Left) BVH$\alpha$ image of M33 taken with SUBARU Suprime-Cam:
 courtesy of V.~Vansevicius, S.~Okamoto, and N.~Arimoto. 
The rectangle represents our mapping area of CO\,({\it J}\,=\,1--0) ($30'\times 20'$).
(Right) Smoothed and clipped integrated intensity
 map of the CO\,({\it J}\,=\,1--0) lines. The contour levels are 1, 2, 4, 6, and $8\K\kms$,
 respectively. \label{fig:field-mom0}}   
\end{figure}

\begin{figure}[p]
\begin{center}
\plotone{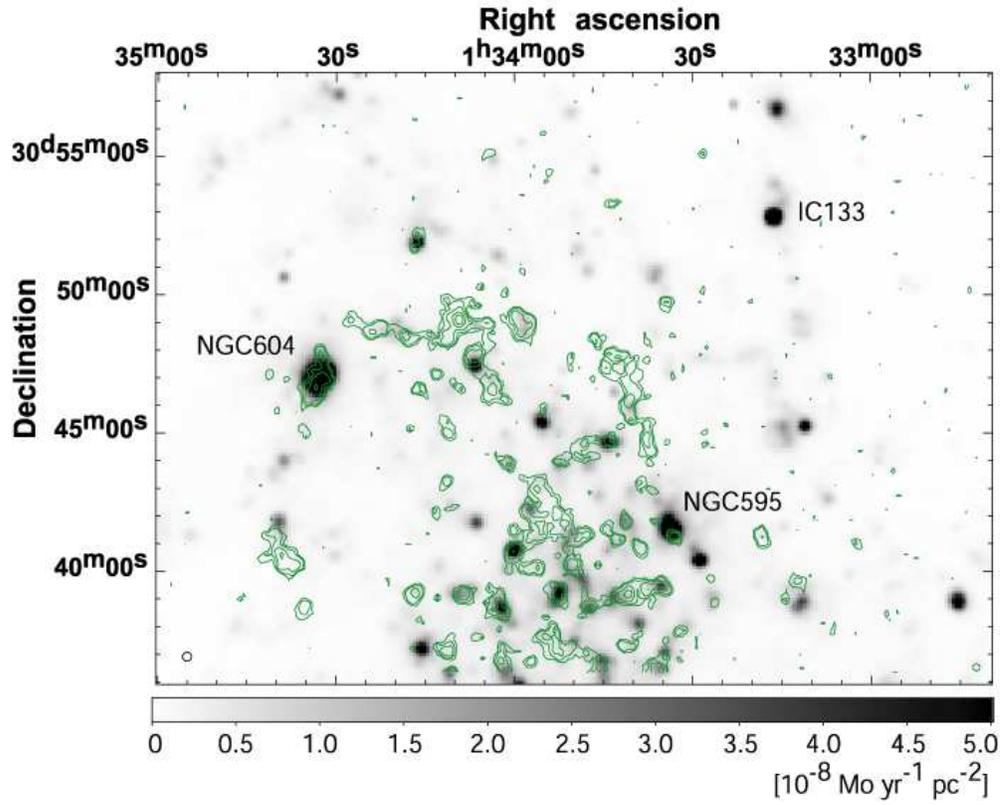}
\end{center}
\caption{CO integrated intensity map (contour) overlaid on the SFR map
convolved into $19''.3$ resolution. The contour levels are 1, 2, 4, 8, and 16 $\K\kms$. The circle at bottom left represents the beam size. \label{fig:imgCOonSFR}}  
\end{figure}

\begin{figure}[p]
\begin{center}
\plotone{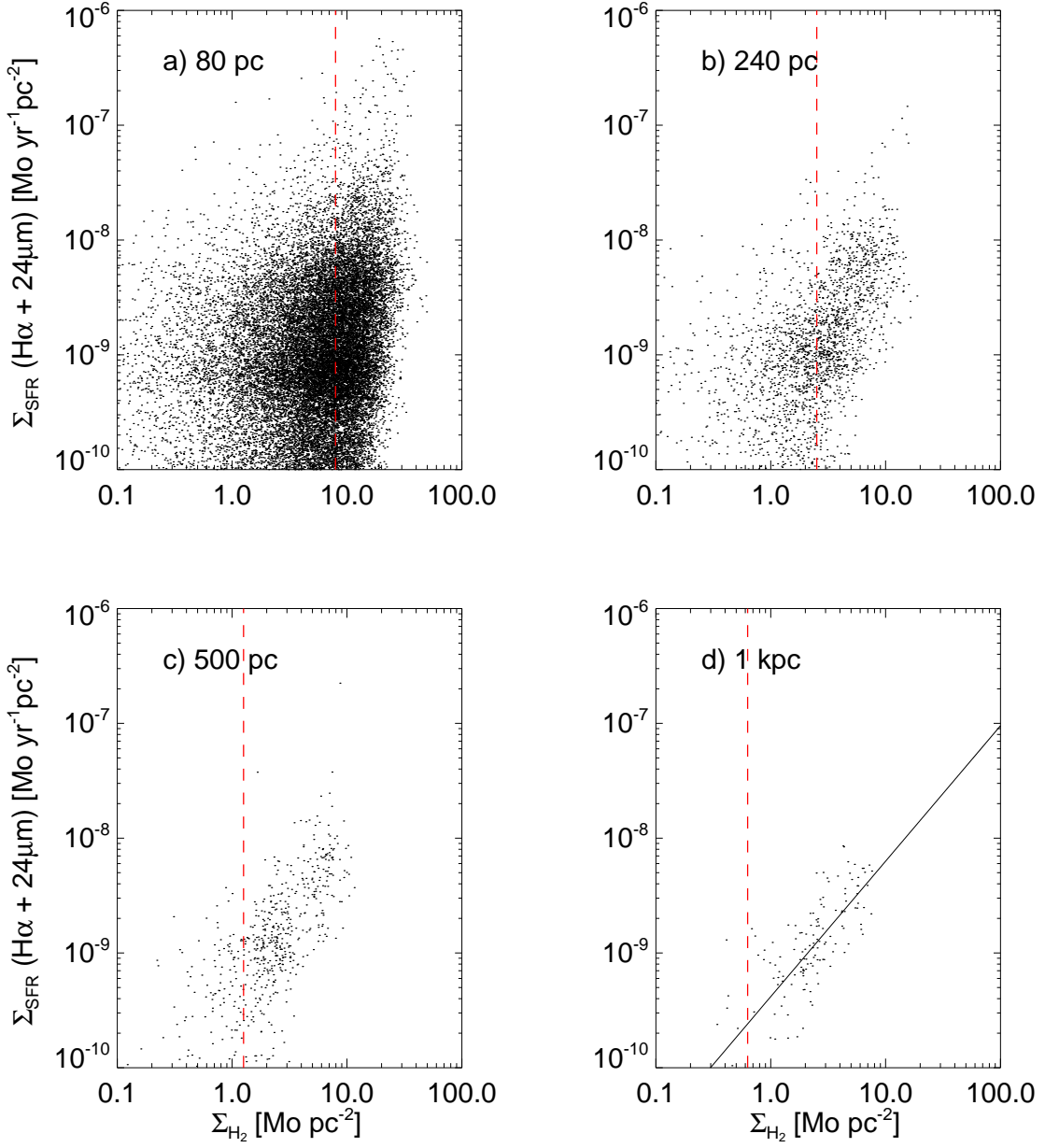}
\end{center}
\caption{Star formation rate per unit area ($\Sigma_{\rm SFR}$) versus surface density of $\rm H_2$ gas ($\Sigma_{\rm H_2}$) for four different resolutions: (a) $19''.3$ ($\sim\!80\pc$), (b) $60''$ ($\sim\!240\pc$), (c) $120''$ ($\sim\!500\pc$), and (d) $240''$ ($\sim\!1\kpc$). The broken red lines in each panel represent $\Sigma_{\rm H_2}=2\sigma$ of the maps. The line in the lower-right panel represents the best least-squares fit to the $\sim\!1\kpc$ resolution data; $\log \Sigma_{\rm SFR}=(1.18\pm 0.11)\log\Sigma_{\rm H_2}-(9.38\pm0.05)$. The data points with $\Sigma_{\rm H_2}<2\sigma$ are not used for the fitting. \label{fig:schmidt}}   
\end{figure}

\end{document}